\documentclass[aps,pra,reprint,superscriptaddress,showpacs]{revtex4-1}
\usepackage{graphicx} 
\usepackage{epstopdf}
\usepackage{color}

\begin{document}


\title{Rydberg excitation of cold atoms inside a hollow core fiber}


\author{Maria Langbecker}
\author{Mohammad Noaman}

\affiliation{Institut f\"{u}r Physik, Johannes Gutenberg-Universit\"{a}t Mainz, 55122 Mainz, Germany}

\author{Niels Kj{\ae}rgaard}
\affiliation{Department of Physics, QSO --- Centre for Quantum Science, and Dodd-Walls Centre for Photonic and Quantum Technologies, University of Otago, Dunedin, New Zealand}

\author{ Fetah Benabid}
\affiliation{GPPMM group, XLIM, UMR 7252 Universit\'{e} de Limoges, 123 Avenue Albert Thomas, 87060 Limoges Cedex, France}

\author{Patrick Windpassinger}
\email[Corresponding author: ]{windpass@uni-mainz.de}
\affiliation{Institut f\"{u}r Physik, Johannes Gutenberg-Universit\"{a}t Mainz, 55122 Mainz, Germany}

\date{\today}

\begin{abstract}
We report on a versatile, highly controllable hybrid cold Rydberg atom fiber interface, based on laser cooled atoms transported into a hollow core Kagom\'{e} crystal fiber. Our experiments are the first to demonstrate the feasibility of exciting cold Rydberg atoms inside a hollow core fiber and we study the influence of the fiber on Rydberg electromagnetically induced transparency (EIT) signals. Using a temporally resolved detection method to distinguish between excitation and loss, we observe two different regimes of the Rydberg excitations: one EIT regime and one regime dominated by atom loss. These results are a substantial advancement towards future use of our system for quantum simulation or information.
\end{abstract}

\pacs{{32.80.Rm}, {03.67.Lx}, {34.35.+a}, {42.50.Gy}}

\maketitle

Rydberg nonlinear quantum optics is a recent and rapidly growing field \cite{Firstenberg2016}. A crucial building block is the strong induced dipole-dipole coupling between two Rydberg atoms which can easily be ten orders of magnitude larger than in conventional ground state systems \cite{Saffman2010}. This is combined with the extraordinary degree of control over the light-matter interactions obtained by electromagnetically induced transparency (EIT) \cite{Fleischhauer2005}. As a consequence, the strong Rydberg nonlinearity yields the possibility of engineering interactions between individual photons \cite{Friedler2005, Gorshkov2011, Pritchard2012}. Effective attractive and repulsive interactions between two photons \cite{Peyronel2012, Firstenberg2013, Dudin2012} and a control of the interaction by microwave fields \cite{Maxwell2013} have been demonstrated. Such a controllable interaction between single photons is ideal for quantum information tools like optical switches, transistors or phase gates \cite{Parigi2012, Firstenberg2013, Dudin2012, Maxwell2014, Gorniaczyk2014, Tiarks2014, Baur2014, Gorniaczyk2016, Tresp2016, Tiarks2016, Saffman2016, Thompson2017}.

Furthermore, controlling the interactions between photons provides the basis for analog photonic quantum simulation \cite{Noh2017}. For example, phase transitions in the Bose-Hubbard model \cite{Huo2012} or relativistic physics \cite{Angelakis2013} could be simulated. It also opens the possibility for investigating the field of many body polariton states \cite{Otterbach2013, Bienias2014, Moos2015, Maghrebi2015}, where a polariton describes a strongly coupled light-matter-system \cite{Fleischhauer2005}. Recently significant theoretical efforts have been devoted to the understanding of the scattering and interaction potentials of Rydberg polaritons \cite{Otterbach2013, Bienias2014, He2014, Moos2015, Maghrebi2015}. 
By spatially confining these polaritons, a strongly interacting one-dimensional system can be created. In such a system, it should for instance be possible to observe crystalline type correlations as known from Tonks-Girardeau gases \cite{Tonks1936, Girardeau1960, Lieb1963, Paredes2004, Kinoshita2004} in a polariton gas \cite{Chang2008, Gorshkov2013, Otterbach2013, Moos2015}. This observation would benchmark photonic quantum simulators.

Rydberg atoms inside hollow core fibers are a promising tool to create strongly interacting one-dimensional many body polariton systems. The first excitation of Rydberg states in a room temperature cesium gas inside a hollow core fiber was reported by Epple et al. \cite{Epple2014}. In a complementary approach, cold atoms transported into a hollow core fiber \cite{Vorrath2010, Bajcsy2011, Blatt2014, Okaba2014} offer important advantages for the initial characterization due to their high controllability. Ground state EIT measurements have already been performed in these systems \cite{DunckerThesis, Blatt2016}. However, when using Rydberg EIT, one major challenge is the understanding and control of the interaction between the strongly polarizable Rydberg states and the fiber walls. Especially the influence of stray electric fields due to adsorbates on the surface has been observed in several experiments \cite{Kuebler2010, Tauschinsky2010, Abel2011, Hattermann2012, Carter2012, Chan2014, Naber2016}. While first attempts have been made to reduce adsorbate fields on quartz surfaces with a specific crystalline structure \cite{Sedlacek2016}, it was strongly debated how much Rydberg excitations inside a hollow core fiber would suffer from surface interactions. In this manuscript, we present for the first time a highly controllable hollow core fiber cold Rydberg atom interface and show how the Rydberg EIT signals inside the fiber are influenced by the fiber.

\begin{figure}
\includegraphics{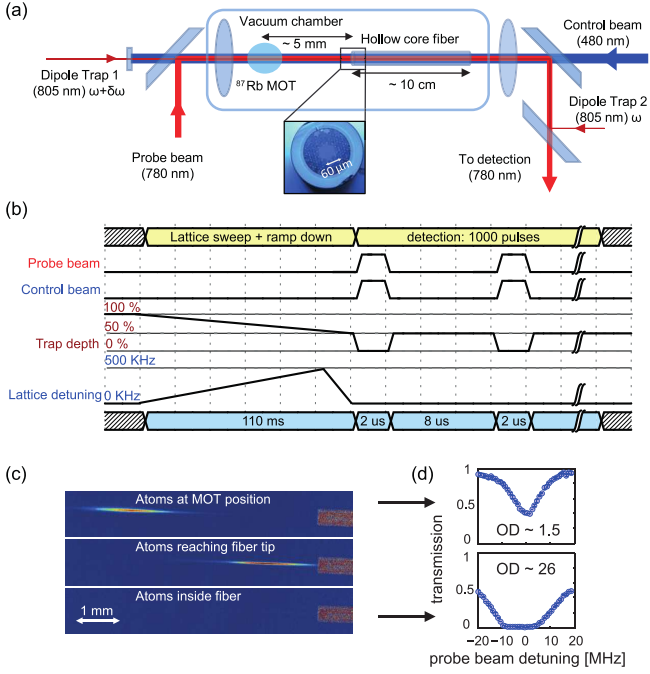}
\caption{\label{fig:setup} (a) Experimental setup. (b) Experimental timing sequence for transport and detection. (c)  Absorption images for the transport process into the fiber. (d) OD measurements inside and outside of the fiber (averaging over 20 repetitions starting from repetition number 1, statistical error bars).}
\end{figure}

Our experimental setup is schematically depicted in Fig. \ref{fig:setup} (a). A 10 cm long hollow-core photonic crystal fiber (HC-PCF, \cite{Couny2007}) with a core diameter of $\rm 60 ~\mu m$ and a mode field diameter of about  $\rm 42 ~\mu m$ is mounted inside a vacuum chamber. We carefully couple all relevant laser beams (two counter-propagating dipole trap beams at 805 nm and two counter-propagating EIT beams at 780 nm and 480 nm) into the fundamental mode of the fiber with more than $90 \%$ coupling efficiency \cite{supplement}. Rubidium 87 atoms are loaded in a magneto-optical trap (MOT) $\sim \rm 5~ mm$ in front of the fiber tip, in which they are cooled to a few tens of microkelvin. The cold atoms are then loaded into a red-detuned optical lattice, created by the two fiber-coupled dipole trap beams \cite{supplement}. 

We use an optical conveyor belt to transport the atoms into the fiber in a highly controlled way \cite{Schrader2001, Okaba2014}. The basic experimental procedure for this is depicted in Fig. \ref{fig:setup} (b). The frequency difference between the two dipole trap beams is increased linearly over 100 ms to detunings up to $\rm 500~kHz$. Simultaneous to this frequency sweep, the laser power is ramped down to $50 \%$ to compensate for increasing trap depth near the fiber tip. Example absorption images of the atoms at different distances from the fiber tip are shown in Fig. \ref{fig:setup} (c), illustrating the transport process. In this way, we transport $5 \times 10^4$ atoms to the tip of the HC-PCF.

As absorption imaging is not possible inside the fiber, we additionally probe the atoms via the absorption of a weak probe beam ($\rm \sim 100~pW$, corresponding to a Rabi frequency $\rm \Omega_p= 2 \pi \, 0.3~MHz$) resonant to the $5 S_{1/2} (F=2) \rightarrow 5 P_{3/2} (F'=3)$ transition, which is coupled through the HC-PCF. During these spectroscopy measurements, we switch off the lattice beams and the $\rm 2~\mu s$ probe pulse is recorded on a photomultiplier tube (PMT). After probing, the atoms are recaptured by switching on the lattice beams again and held in the lattice for $\rm \sim 8~\mu s$, as shown in  Fig. \ref{fig:setup} (b). This whole sequence of $\rm10~\mu s$ is repeated for 1000 times. To obtain the absorption spectrum, we keep the probe beam at a fixed frequency for each experimental run and only change this frequency between runs \cite{supplement}. For each spectrum, the optical depth (OD) is calculated by a fit to the experimentally determined transmission $T$ of the probe pulse: 
\begin{equation}
T = \exp\left(-\frac{OD}{1+4\left(\Delta/\gamma\right)^2}\right),
\label{eq:OD}
\end{equation}
where $\Delta$ is the detuning of the probe beam and $\gamma$ the natural linewidth.

Two such absorption spectra are shown in Fig. \ref{fig:setup} (d), one for the atoms at the MOT position, the second for atoms about $\rm5~mm$ inside the fiber. As the atoms are transported into the fiber, the optical depth increases. This corresponds well to theoretical expectations as the probe beam is focused towards the fibertip \cite{Bajcsy2011,supplement}. Once the atoms have entered the hollow core fiber, we observe that the optical depth and thus the atom number stays constant. Lifetimes and temperatures inside the fiber (between $\rm 120 ~ms$ and $\rm 180 ~ms$ and $\rm 500 ~\mu K$) are determined with a release-and-recapture measurement and are comparable to the measurements outside the the fiber tip ($\rm 200 ~ms$ and $\rm 300 ~\mu K$) using a standard absorption imaging technique \cite{supplement}. In both cases, the numbers are limited by increased scattering of our near-detuned dipole trap close to the fiber tip. This shows that we can control the transport process into and inside the fiber and that the fiber has only moderate influence on atom numbers and temperatures.

\begin{figure}
\includegraphics{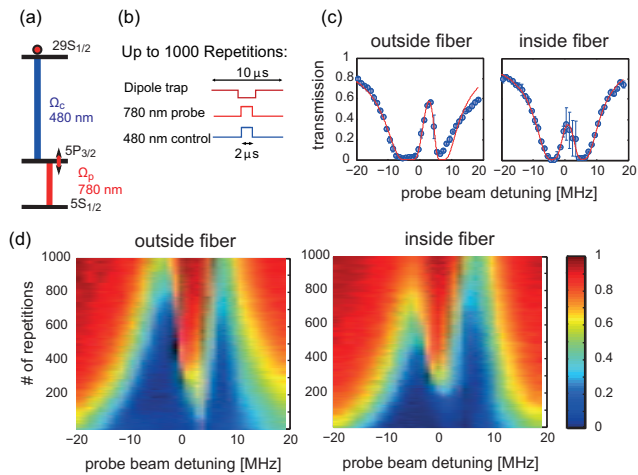}
\caption{\label{fig:eit_distance} (a) Rydberg excitation scheme. (b) Single-pulse measurement sequence. (c) EIT signals inside and outside of the fiber (\textcolor{blue}{$\circ$}, averaging over 20 repetitions starting from repetition number 201, statistical error bars) and EIT fit (solid line). (d) Time resolved EIT signals inside and outside of the fiber (Moving average over 20 neighboring repetitions).}
\end{figure}

We exploit a resonant two photon ladder EIT scheme to excite and detect Rydberg atoms, as sketched in Fig. \ref{fig:eit_distance} (a). For this, we keep the control beam at $\rm 480~nm$ at resonance to the $5 P_{3/2} (F'=3) \rightarrow 29 S_{1/2} (F''=\text{unresolved})$ Rydberg transition using an EIT locking scheme \cite{Abel2009}, while we scan the $\rm 780~nm$ probe laser over the $5 S_{1/2} (F=2) \rightarrow 5 P_{3/2} (F'=3)$ resonance. Both laser line widths are on the order of $\rm100 ~kHz$. We switch control and probe beam on and off simultaneously as shown in Fig. \ref{fig:eit_distance} (b) and overlap both beams spatially by coupling them into the fundamental mode of the hollow core fiber. The absorption spectrum is obtained in the same way as for the OD measurements. For further analysis, we fit the spectra with an EIT formula according to Ref. \cite{Tiarks2016}, which is valid in our limit of low probe Rabi frequency:
\begin{equation}
T = \exp{\left[ -OD \, \text{Im} \left( \chi \right) \right]}
\label{eq:EIT}
\end{equation}
with the susceptibility
\begin{equation}
\chi = i \gamma \left( \gamma - 2i \Delta + \frac{\left|\Omega_c \right|^2}{\gamma_{\text{ryd}} -2i\left( \Delta_c + \Delta \right) } \right)^{-1},
\end{equation}
where the detuning and decay rates of the probe and the control transition are denoted by $\Delta$ and $\gamma$ and by $\Delta_c $ and $\gamma_{\text{ryd}}$ respectively. For the Rydberg state, we are interested in the additional decay or dephasing rate $\gamma_{\text{ryd,2}}=\gamma_{\text{ryd}}-\gamma_{\text{29S}}$, where $\gamma_{\text{29S}}=2\pi / (\rm 21.7 ~ \mu s)$ is the natural linewidth \cite{sibalic2016}. Additional dephasing can for example stem from inhomogeneous electric or magnetic fields. $\Omega_c$ is the Rabi frequency of the control beam. In the absence of a control beam, i.e. in the case of $\left|\Omega_c \right| = 0$, eq. (\ref{eq:EIT}) reduces to eq. (\ref{eq:OD}).

Figure \ref{fig:eit_distance} (c) shows typical EIT signals inside and outside of the fiber. As the frequency of the probe beam is changed, the EIT window opens up at the two-photon resonance position. We observe a maximum transmission of about $60 \%$ outside of the fiber and of about $40 \%$ inside the fiber. The asymmetric outer part of the signal outside the fiber can also be observed in OD only measurements and in our opinion does not play an important role regarding the EIT process. By fitting eq. (\ref{eq:EIT}) to our experimental data, we determine the control Rabi frequency to be $2\pi \,  \rm(9.9 \pm 0.8)~MHz$ outside and $2\pi \,  \rm(9.5 \pm 0.6)~MHz$ inside the fiber, corresponding to control beam powers of about 2 mW, which corresponds well to our measured powers taking into account losses, e.g. due to uncoated vacuum windows. To extract the influence of the fiber on the lifetime of the Rydberg state, we compare the decay rates $2\pi \, \rm(2.6  \pm 0.75)~MHz$ inside and $2\pi \, \rm(0.9  \pm 0.4)~MHz$ outside the fiber. The outside value is typical for cold atom Rydberg experiments \cite[e.g.][]{Tiarks2016}, whereas the increase inside the fiber hints at interactions of the Rydberg atoms with the fiber walls. This assumption is confirmed by a relative control beam shift between the measurement inside and outside of the fiber of $\rm(2.2 \pm 1)~MHz$, which we determine from the detunings $\Delta$ and $\Delta_c$. Given the polarizability of the  $29 S_{1/2}$ state of $\alpha = \rm 1.14 ~MHz~ cm^2 / V^2$ \cite{sibalic2016}, this shift would correspond to an electric field of $\rm(2 \pm 1.3)~V/cm$. This is comparable to electric fields due to adatoms on dielectric surfaces \cite{Kuebler2010, McGuirk2004} and small compared to fields due to adatoms on metallic or coated metallic surfaces \cite{Hattermann2012, Naber2016}. In our case, the adatom distribution can be inhomogeneous along the fiber axis. We note that the shift is small with the respect to the EIT linewidth and stays constant between different measurement days, suggesting that an adatom saturation of the surface has already been reached.

To gain further insight into the interaction processes of the Rydberg atoms with the fiber, we perform a time-resolved analysis of our data. Instead of showing an averaged signal, Fig. \ref{fig:eit_distance} (d) presents EIT signals inside and outside of the fiber as function of the measurement repetition number. Apart from the shift between the EIT peaks inside and outside the fiber as already discussed for Fig. \ref{fig:eit_distance} (c), we observe two additional features. First, the OD decreases during the measurement sequence, from 32 to 2 outside and from 19 to 1 inside of the fiber, which indicates a loss of atoms. Qualitatively, this behaviour can also be observed in OD only measurements and is related to our measurement method rather than the lifetime of atoms in the lattice. We will discuss the explicit influence of the EIT process on this OD decrease in the following. Second, a shifting of the EIT peak, i.e. the 2 photon resonance position, with the number of repetitions is noticeable. Density-related effects \cite{Han2016} could explain shifts as the atomic density decreases during the measurement sequence, but our calculations show that they are small for our experimental parameters. However, we find a similar shift when we compare coherent excitation with atom loss, as will be discussed in the following. In conclusion, we also note that both additional effects occur both inside and outside the fiber and thus the fiber does not influence our signal significantly.  

\begin{figure}
\includegraphics{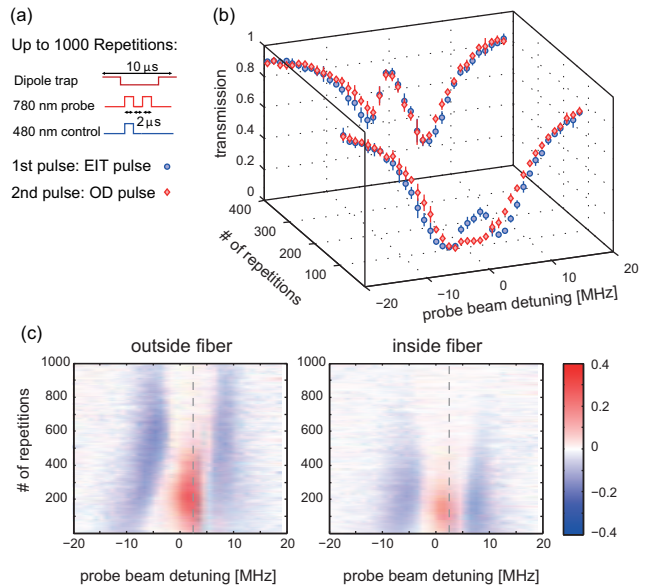}
\caption{\label{fig:eit_time} (a) Two-pulse measurements sequence. (b) Comparison of OD (\textcolor{red}{$\diamond$}) and EIT (\textcolor{blue}{$\circ$}) pulses inside the fiber for different number of repetitions (averaging over 20 repetitions, statistical error bars). (c) Time resolved difference between EIT and OD pulses (Moving average over 20 neighboring repetitions). The dashed gray line marks the cut shown in Fig. \ref{fig:eit_time_loss}.}
\end{figure}

In order to further investigate the loss dependent processes and to determine whether we see a coherent EIT signal or mainly a loss of atoms after excitation into the Rydberg state, we directly probe the atom loss in a temporally resolved way. To this end, we add another probe pulse directly after each two photon EIT measurement as shown in Fig. \ref{fig:eit_time} (a). During the first pulse, both the control laser and the probe laser are switched on and the EIT signal is recorded. In the second OD only pulse, the control laser remains switched off. The difference between the two probe pulses is a direct measure of the atom loss induced by the Rydberg EIT process.

In these measurements, we can clearly distinguish between two different time regimes. Fig. \ref{fig:eit_time} (b) shows both EIT and OD signals inside the fiber for two representative repetition numbers. For early repetition numbers in the measurement sequence, the EIT and the OD signals are clearly distinct. The EIT pulse shows the transparency window opening up, which is absent in the OD pulse. For larger repetition numbers, both signals become equal as the OD pulse also develops a peak structure. This can be attributed to atom loss, e.g. due to excitation and ionization of Rydberg atoms or due to loss to other Rydberg states. Note that this loss peak in both OD and EIT signal is shifted about $\rm 2.5~MHz$ with respect to the EIT peak appearing for earlier repetitions. This shift is consistent with the one observed in Fig. \ref{fig:eit_distance} (d) and marks the transition from coherent EIT regime to atom loss dominated regime. A similar shift is observable in the measurements outside the fiber.

To visualize the temporal evolution, we subtract the two-photon EIT and the OD measurement (i.e. the two datasets in each of the two plots of Fig. \ref{fig:eit_time} (b)) and plot this difference as a function of the repetition number in  Fig. \ref{fig:eit_time} (c), presenting both inside and outside the fiber measurements. These results confirm the previous assumption of two different time regimes. In both cases, a clear EIT signal is visible at early times, confirming a coherent excitation. Inside the fiber, this EIT peak vanishes after about 300 pulses, which indicates that now both EIT and OD pulse have become equal. Since only loss processes can also be observed in the OD pulse, they have become more dominant than the coherent EIT process for these later times. Outside of the fiber, we see a qualitatively similar behavior, but a clear EIT signal is still visible for repetition numbers up to 600. That suggests that loss processes are enhanced by the fiber and the loss-dominated regime starts earlier inside the fiber. This meets our expectations as inside the fiber atoms are lost once they hit the fiber wall, while outside the fiber they can still contribute to the signal even after expansion beyond the beam waist. Further, we notice that in this coherent EIT signal no shift with increasing number of repetitions occurs. This confirms our assumption that the previously observed time-dependent shift is due to loss processes.

\begin{figure}
\includegraphics{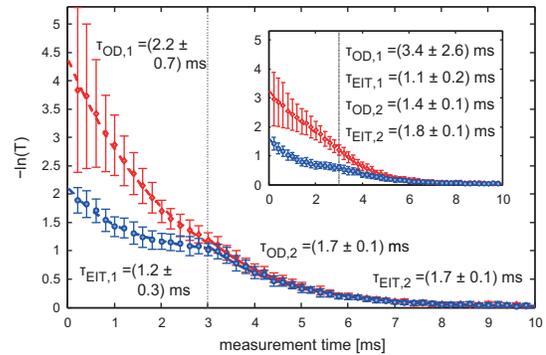}
\caption{\label{fig:eit_time_loss} Comparison of OD (\textcolor{red}{$\diamond$}) and EIT (\textcolor{blue}{$\circ$}) signals at $\Delta = 2.5 \, \text{MHz}$ as function of measurement time (averaging over 20 repetitions, statistical error bars). The main figure shows the signals inside the fiber, the inset the signals outside of the fiber. The dashed (solid) lines show an exponential fit to data below (above) 300 repetitions (time scale 1 (2)). The two regimes are separated by a dotted gray line. The decay times for each regime and each dataset are given in the figure (errors from the fit).}
\end{figure}

To determine the time scales for the two different regimes, we make a vertical cut through Fig. \ref{fig:eit_time} (c) at the position of the initial EIT peak at $\Delta = \rm 2.5~MHz$, marked by the dashed gray line. The logarithm of the transmission for OD and EIT pulse is plotted as function of measurement time in Fig. \ref{fig:eit_time_loss}, inside (main plot) and outside (inset) of the fiber. Inside the fiber, there exist two distinct time scales with a sharp cut at around 3 ms, i.e. 300 repetitions. The transition between the two time regimes seems to happen when the magnitudes of OD and EIT pulse become the same. Outside the fiber, this transition is less distinct, although the EIT signal also shows a different behavior before and after 3 ms. We have fitted exponential loss curves to all datasets and give the decay rates in Fig. \ref{fig:eit_time_loss}, together with their respective errors from the fit. Only the initial loss of atoms (OD time scale 1) depends on whether the atoms are outside or inside of the fiber, with a faster loss inside the fiber. However, we cannot make a quantitative statement due to the large error for this time scale. The EIT decay rates in both regimes as well as the later OD decay rate are not influenced by the fiber. Thus, while we do observe an accelerated overall loss of atoms due to Rydberg excitations inside the fiber, for time scales up to a few ms the fiber has no significant influence on the occurrence of the EIT signal itself.

In conclusion, we have demonstrated the first Rydberg excitations of cold atoms inside a hollow core fiber. We are able to produce a highly controllable sample of atoms and we find our system well suited as a hybrid cold Rydberg atom fiber interface. We have studied the influence of the fiber on the EIT signal and we observe both a broadening and shift of our signals on the order of their linewidth. Only at positions very close to the fiber tip, we see such a strong influence that we have not been able to produce a clear EIT signal, possibly due to large inhomogeneous electric fields due to adatoms as also observed in numerous other experiments \cite{Kuebler2010, Tauschinsky2010, Abel2011, Hattermann2012, Carter2012, Chan2014, Naber2016}. We are currently further investigating the influences of different types of fibers on our EIT signal in a separate setup and are testing possible techniques to overcome these interaction effects, e.g. by coating of the inner fiber core or by light-induced atomic desorption (LIAD).
The effects of the fiber on coherent Rydberg excitation were further quantified through a time resolved detection method. Here, we have found two different regimes of Rydberg excitations to exist: one EIT regime and one regime dominated by atom loss. As within the EIT regime the fiber does not have a significant influence on the occurrence of the EIT signal, we believe that our system is an important step towards future use of hybrid systems for quantum simulation or information. One further possible future application of our system is to study the propagation of excitations and correlations in dense extended one-dimensional media \cite{Schempp2015, Jennewein2016, Bromley2016, Ruostekoski2016}.

\begin{acknowledgments}
We thank Klaus Sengstock and Hannes Duncker for providing the basic experimental setup and the preliminary work to make these results possible. We thank Ximeng Zheng, Beno\^{i}t Debord and Fr\'{e}d\'{e}ric G\'{e}r\^{o}me for design and production of the hollow core fibers for our experiment. We gratefully acknowledge financial support by the Stufe 1 Funding of the University of Mainz, FP7-PEOPLE-2012-ITN-317485 (Qtea) and the DFG Priority Program SPP 1929 (GiRyd).
\end{acknowledgments}

%

\end{document}